\documentclass{pnastwo}

\usepackage{graphicx}
\usepackage{times}

\begin{document}



\conflictofinterest{Conflict of interest footnote placeholder}

\track{This paper was submitted directly to the PNAS office.}




\title{Propagation of large concentration changes in reversible protein binding networks}




\title{Propagation of large concentration changes in reversible protein binding networks}
\author{Sergei Maslov \affil{1}{Department of Condensed Matter Physics
and Materials Science, Brookhaven National Laboratory, Upton, New
York 11973, USA}
\thanks{Corresponding author, email:
maslov@bnl.gov}, I. Ispolatov \affil{2}{Ariadne Genomics, Inc.,
9430 Key West Ave, Suite 113, Rockville, Maryland 20850, USA}
\thanks{Corresponding author, email: slava@ariadnegenomics.com}
}
\contributor{Submitted to Proceedings of the National Academy of Sciences
of the United States of America}
\maketitle

\begin{article}

\begin{abstract}
We study how the dynamic equilibrium of the reversible
protein-protein binding network in yeast {\it S. cerevisiae}
responds to large changes in abundances of individual proteins.
The magnitude of shifts between free and bound concentrations of
their immediate and more distant neighbors in the network is
influenced by such factors as the network topology, the
distribution of protein concentrations among its nodes, and the
average binding strength. Our primary conclusion is that on
average the effects of a perturbation are strongly localized and
exponentially decay with the network distance away from the
perturbed node. This explains why, despite globally connected
topology, individual functional modules in such networks are able
to operate fairly independently. We also found that under specific
favorable conditions, realized in a significant number of paths in
the yeast network, concentration perturbations can selectively
propagate over considerable network distances (up to four steps).
Such "action-at-a-distance" requires high concentrations of
heterodimers along the path as well as low free (unbound)
concentration of intermediate proteins.
\end{abstract}




\section*{Introduction}
Recent high-throughput experiments performed in a wide variety of
organisms revealed networks of protein-protein physical
interactions (PPI) that are interconnected on a genome-wide scale.
In such ``small-world'' PPI networks most pairs of nodes can be
linked to each other by relatively short chains of interactions
involving just a few intermediate proteins \cite{wagner2001ypi}.
While globally connected architecture facilitates biological
signaling and possibly ensures a robust functioning of the cell
following a random failure of its components \cite{jeong2001lac},
it also presents a potential problem by providing a conduit for
propagation of undesirable cross-talk between individual
functional modules and pathways. Indeed, large (several-fold)
changes in proteins' levels in the course of activation or
repression of a certain functional module affect bound
concentrations of their immediate interaction partners. These
changes have a potential to cascade down a small-world PPI network
affecting the equilibrium between bound and unbound concentrations
of progressively more distant neighbors including those in other
functional modules. Most often such indiscriminate propagation
would represent an undesirable effect which has to be either
tolerated or corrected by the cell. On the other hand, a
controlled transduction of reversible concentration changes along
specific conduits may be used for biologically meaningful
signaling and regulation. A routine and well known example of such
regulation is inactivation of a protein by sequestration with its
strong binding partner.

In this study we quantitatively investigate  how large
concentration changes propagate in the PPI network of yeast {\it
S. cerevisiae}.
We focus on the non-catalytic or reversible binding interactions
whose equilibrium is governed by the Law of Mass Action (LMA) and
do not consider irreversible, catalytic processes such as protein
phosphorylation and dephosphorylation, proteolytic cleavage, etc.
While such catalytic interactions constitute the most common and
best studied mechanism of intracellular signaling, they represent
only a rather small minority of all protein-protein physical
interactions (for example, only $\sim$5\% links in the yeast
network used in our study involve a kinase).

Furthermore, the balance between free and bound concentrations of
proteins matters even for irreversible (catalytic) interactions.
For example, the rate of a phosphorylation reaction
depends on the availability of free kinases and substrate proteins
which are both controlled by the LMA equilibrium calculated here.
Thus perturbations of equilibrium concentrations considered in
this study could be spread even further by other mechanisms such
as transcriptional and translational regulation, and irreversible
posttranslational protein modifications.

\section*{Results}
To illustrate general principles on a concrete example, in this
study we used a highly curated genome-wide network of
protein-protein physical interactions in yeast ({\it S.
cerevisiae}), which, according to the BIOGRID database
\cite{BIOGRID}, were independently confirmed in at least two
publications. We combined this network with a genome-wide dataset
of protein abundances in the log-phase growth in rich medium,
measured by the TAP-tagged western blot technique
\cite{ghaemmaghami2003gap}. Average protein concentrations in this
dataset range between 50 and 1,000,000 molecules/cell with the
median value around 3000 molecules/cell. After keeping only the
interactions between proteins with known concentrations we were
left with 4185 binding interactions among 1740 proteins (Table S1).
The BIOGRID database \cite{BIOGRID} lists all
interactions as pairwise and thus lacks information about
multi-protein complexes larger than dimers. Thus in the main part
of this study we consider only homo- and hetero-dimers and ignore
the formation of higher-order complexes. In the Supplementary
materials we show that the reliable data on multi-protein
complexes can be easily incorporated into our analysis.
Furthermore, we demonstrate that taking into account such
complexes leaves our results virtually unchanged (see
supplementary Table S4 and Fig. S3).

The state of the art genome-wide PPI datasets lack information on
dissociation constants $K_{ij}$ of individual interactions. The
only implicit assumption is that the binding is sufficiently
strong to be detectable by a particular experimental technique
(some tentative bounds on dissociation constants detectable by
different techniques were reported recently
\cite{piehler2005nmm}). A rough estimate of the average binding
strength in functional protein-protein interactions could be
obtained from the PINT database \cite{PINT}. This database
contains about 400  experimentally measured dissociation constants
between wildtype proteins from a variety of organisms. In
agreement with predictions of Refs.
\cite{lancet1993pmm,deeds2006spm} the histogram of these
dissociation constants has an approximately log-normal shape. The
average relevant for our calculations is that of the {\it
association} constant $\langle 1/K_{ij} \rangle=$1/(5nM). Common
sense dictates that the dissociation constant of a functional
binding between a pair of proteins should increase with their
abundances. The majority of specific physical interactions between
proteins are neither too weak (to ensure a considerable number of
bound complexes) nor unnecessarily strong. Indeed, there is little
evolutionary sense in increasing the binding strength between a
pair of proteins beyond the point when both proteins (or at least
the rate limiting one) spend most of their time in the bound
state. The balance between these two opposing requirements is
achieved by the value of dissociation constant $K_{ij}$ equal to a
fixed fraction of the largest of the two abundances $C_i$ and
$C_j$ of interacting proteins. In our simulations
we used $K_{ij}=\max(C_i,C_j)/20$ in which case the average
association constant nicely agrees with its empirical value
(1/(5nM)) observed in the PINT database \cite{PINT}.
In addition to this, perhaps, more realistic assignment of
dissociation constants we also simulated binding networks in which
dissociation constants of all 4185 edges in our network are {\it
equal to each other} and given by 1nM, 10nM, 100nM, and 1$\mu$M.

\subsection*{Numerical calculation of bound and free (unbound) equilibrium
concentrations.} The Law of Mass Action (LMA) relates the free
(unbound) concentration $F_i$ of a protein to its total (bound and
unbound) concentration $C_i$ as
\begin{equation}
F_i=\frac{C_i}{1+\sum_{j} F_j/K_{ij}} \qquad .
\label{eq1}
\end{equation}
Here the sum is over all specific binding partners of the protein
$i$ with free concentrations $F_j$ and dissociation constants
$K_{ij}$. While in the general case these nonlinear equations do
not allow for an analytical solution for $F_i$, they are readily
solved numerically e.g. by successive iterations.

\subsection*{Concentration-coupled proteins.}
To investigate how large changes in abundances of individual protein
affect the equilibrium throughout the PPI network we performed a
systematic numerical study in which we recalculated the equilibrium free
concentrations of all protein nodes following a twofold increase in
the total concentration of just one of them: $C_i \to 2C_i$. This
was repeated for the source of twofold perturbation spanning the set of
all 1740 of proteins in our network \cite{linear_response}.
The magnitude of the initial perturbation was
selected to be representative of a typical shift in gene
expression levels or protein abundances following a change in
external or internal conditions. Thus here we simulate the
propagation of functionally relevant changes in protein
concentrations and not that of background stochastic fluctuations.
A change in the free concentration $F_j$ of another protein was
deemed to be significant if it exceeded the 20\% level, which
according to Ref. \cite{newman2006scp} is the average magnitude of
cell-to-cell variability of protein abundances in yeast. We refer
to such protein pairs $i \to j$ as {\it concentration-coupled}.
The detection threshold could be raised simultaneously with the
magnitude of the initial perturbation. For example, we found that
the list of concentration-coupled pairs changes very little if
instead of twofold (+100\%) perturbation and the 20\% detection
threshold one applies a sixfold (+500\%) initial perturbation and
twofold (100\%) detection threshold.

In general we found that lists of concentration-coupled proteins
calculated for different assignments of dissociation constants
strongly overlap with each other. For example, more than 80\% of
concentration-coupled pairs
observed for the variable $K_{ij}=\max(C_i,C_j)/20$ assignment
described above were also detected for the uniform assignment
$K_{ij}=\mathrm{const}=10$nM (for more details see the
supplementary table S3) This relative robustness of our results
allowed us to use the latter conceptually simplest case
to illustrate our findings in the rest of the manuscript.

The complete list of concentration-coupled pairs is included in
the supplementary materials. Given the incompleteness and uncertainty
in our knowledge of the network topology, protein abundances,  and
values of dissociation constants, these lists provide only a rough
estimate of the actual magnitude of perturbations that could be
measured experimentally.

\subsection*{Central observations.}
We found that:
\begin{itemize}
\item {\it On average}, the magnitude of cascading changes in equilibrium
free concentrations exponentially decays with the distance from
the source of a perturbation. This explains why, despite a
globally connected topology, individual modules in such networks
are able to function fairly independently.
\item Nevertheless,
specific favorable conditions identified in our study cause
perturbations to selectively affect proteins at considerable
network distances (sometimes as far as four steps away from the
source). This indicates that in general, such cascading changes
{\it could not be neglected} when evaluating the consequences of
systematic changes in protein levels, e.g. in response to
environmental factors, or in gene knockout experiments. Conditions
favorable for propagation of perturbations combine high yet
monotonically decreasing concentrations of all heterodimers along
the path with low free (unbound) concentrations of intermediate
proteins. While reversible protein binding links are symmetric,
the propagation of concentration changes is usually asymmetric
with the preferential direction pointing down the gradient in the
total concentrations of proteins.
\end{itemize}

\subsection*{Examples of multi-step cascading changes.}
\begin{figure}
\includegraphics[width=.4\textwidth]{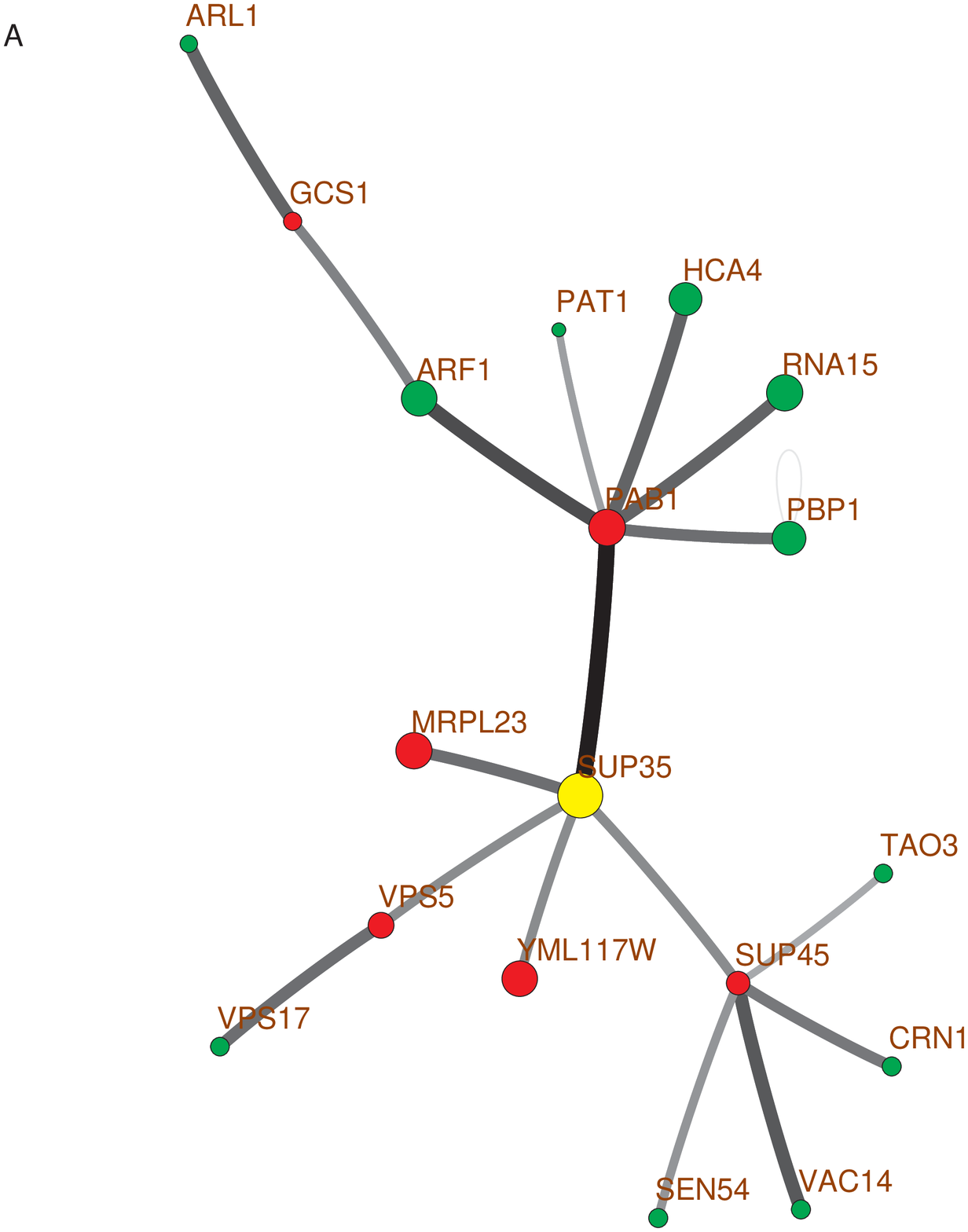}
\includegraphics[width=.4\textwidth]{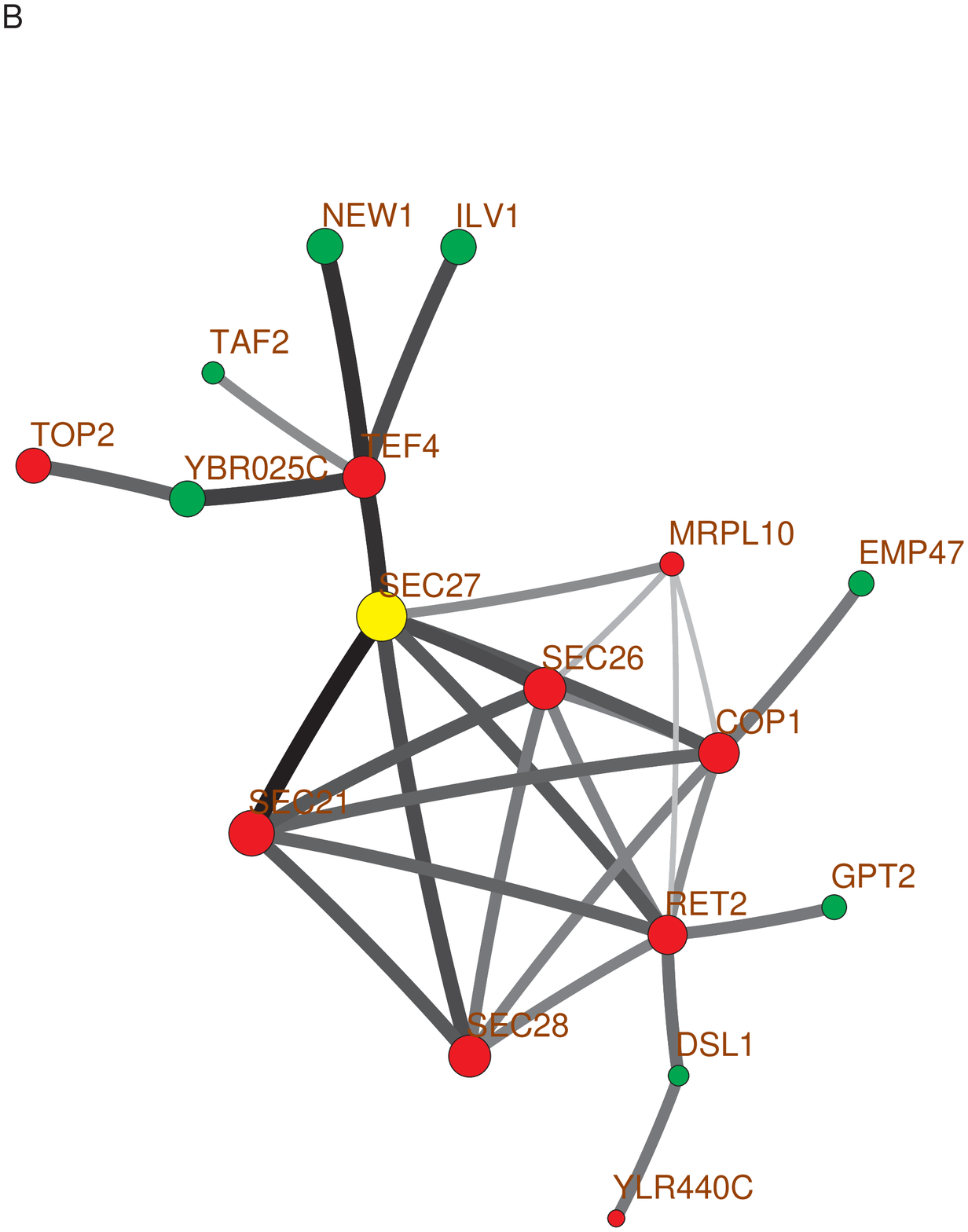}
\caption{
\label{fig_1}
Two cases of propagation of large concentration changes
in the yeast protein binding network.
The total (bound + unbound) concentration of the protein marked with the yellow circle
(the SUP35 protein (A), the SEC27 protein (B))
was increased twofold from its wildtype value in the rich growth medium
\cite{ghaemmaghami2003gap}.
Red and green circles mark all other proteins whose equilibrium free (unbound)
concentrations have increased (green) or decreased (red) by more than 20\%.
The area of each circle is proportional to the logarithm of
the change in free concentration.
Edges show all physical interactions among this group of proteins with
the shade of gray proportional to the logarithm of the
equilibrium concentration of the corresponding dimer calculated for
$K_{ij}=\mathrm{const}=10$nM.
}
\end{figure}

In Fig.~\ref{fig_1}AB we illustrate these
observations using two examples.
In each of these cases the twofold increase in the abundance of just one
protein (marked with the yellow circle in the center of each panel)
has significantly ($>$ 20\%) affected
equilibrium free concentrations
of a whole cluster of proteins some as far as 4 steps
away from the source of the perturbation.
However, the propagation beyond
immediate neighbors is rather specific.
For example, in the case of
SUP35 (Fig.~\ref{fig_1}A)
only 1 out of 169 of its third
nearest neighbors were affected above the 20\% level.
Note that changes in free concentrations
generally sign-alternate with the network distance from the source.
Indeed, free concentrations of immediate binding partners of the
perturbed protein usually drop as more of them become bound
in heterodimers with it. This, in turn,
lowers concentrations of the next-nearest heterodimers
and thus {\it increases} free concentrations of proteins at distance 2 from the
source of perturbation, and so on.

\subsection*{Exponential decay with the network distance.}
\begin{figure}
\includegraphics[width=.95\columnwidth]{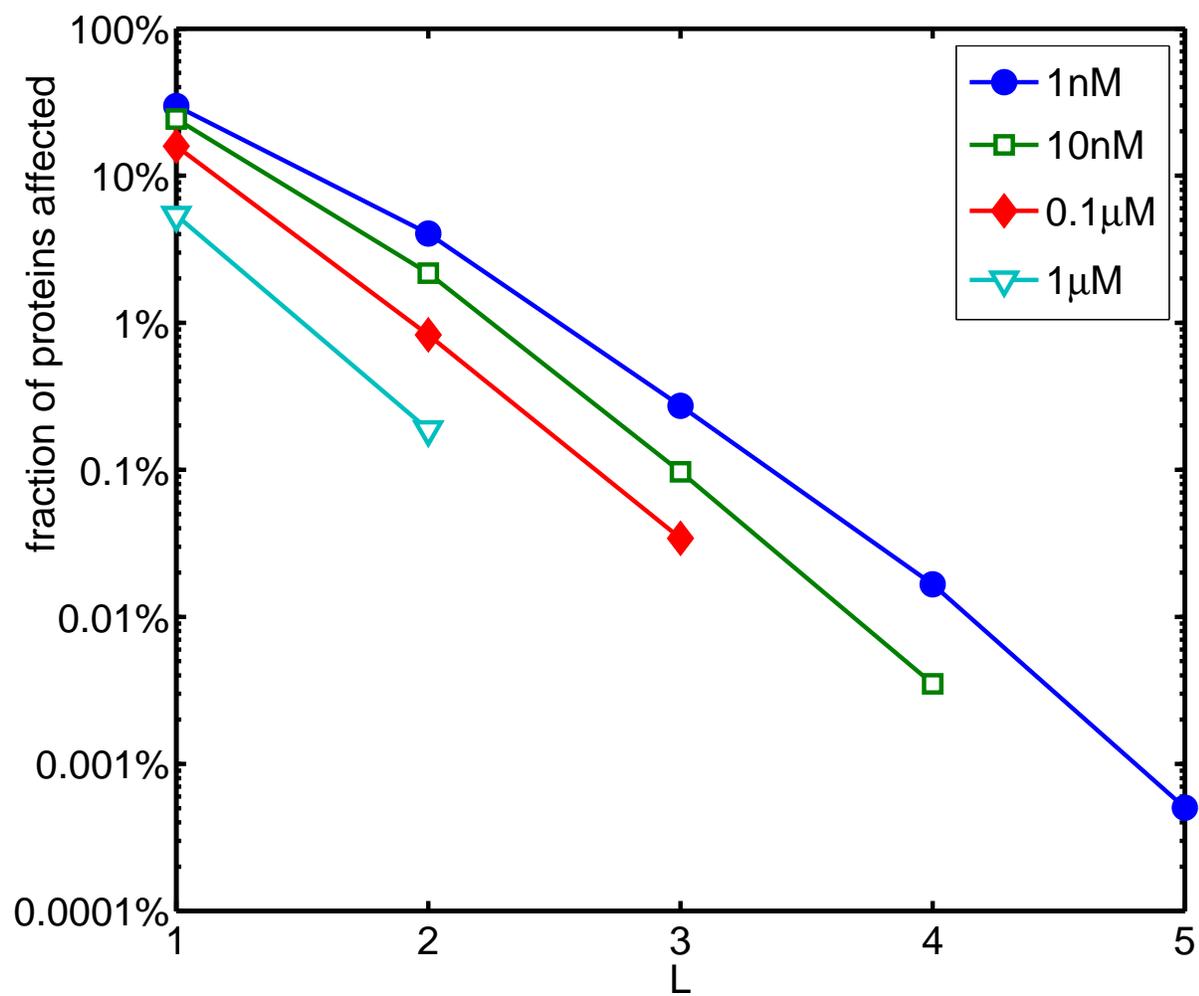}
\caption{ \label{fig_2}
Indiscriminate propagation of concentration perturbations
is exponentially suppressed. The
fraction of proteins with free concentrations affected by more than
20\% among all proteins at network  distance $L$ from the
perturbed protein. Different curves correspond to simulations
with $K_{ij}=\mathrm{const}=1$nM
(solid circles), $10$nM (empty squares),
$0.1\mu$M (solid diamonds), and $1\mu$M (empty triangles). }
\end{figure}

The results of our quantitative network-wide analysis of these
effects are summarized in Fig.~\ref{fig_2} and Table~\ref{tab1}.
From  Fig ~\ref{fig_2}
one concludes that the fraction of proteins
with significantly affected free concentrations rapidly
(exponentially) decays with the length $L$ of the shortest path
(network distance) from the perturbed protein. The same statement
holds true for bound concentrations if the distance is measured as
the shortest path from the perturbed protein to any of the two
proteins forming a heterodimer. Thus, on average, the propagation
of concentration changes along the PPI network is indeed
considerably dampened.
On the other hand, from
Table~\ref{tab1}
one concludes that the total number of multi-step chains along
which concentration changes propagate with little attenuation
remains significant for all but the largest values of the
dissociation constant.
These two observations do not contradict each other since the
number of proteins separated by distance $L$  (the last column in
Table~\ref{tab1}) rapidly grows with $L$.

\subsection*{Conditions favoring the multi-step propagation of
perturbations.}
\begin{figure}[ht]
\includegraphics[width=.3\columnwidth]{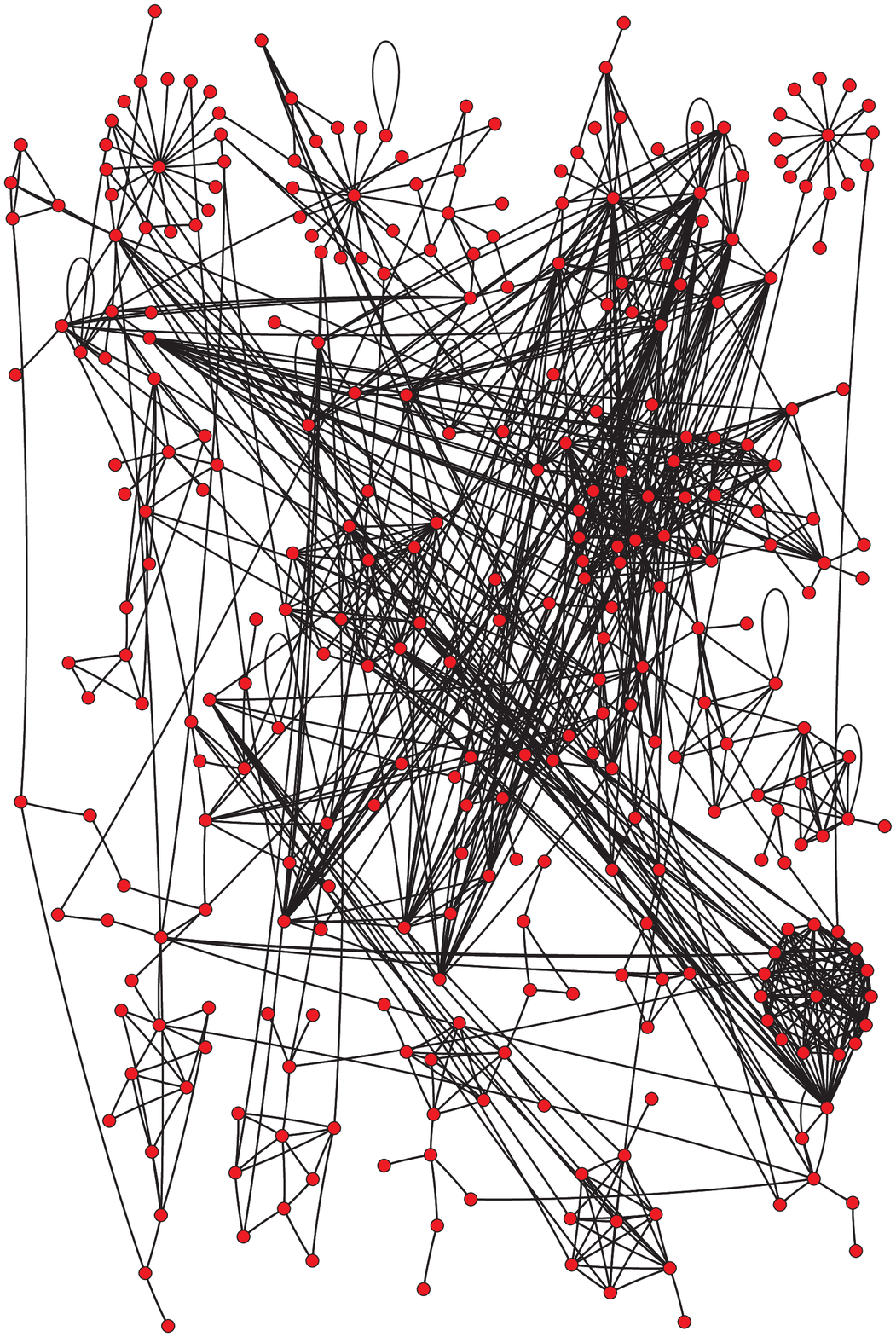}
\includegraphics[width=.3\columnwidth]{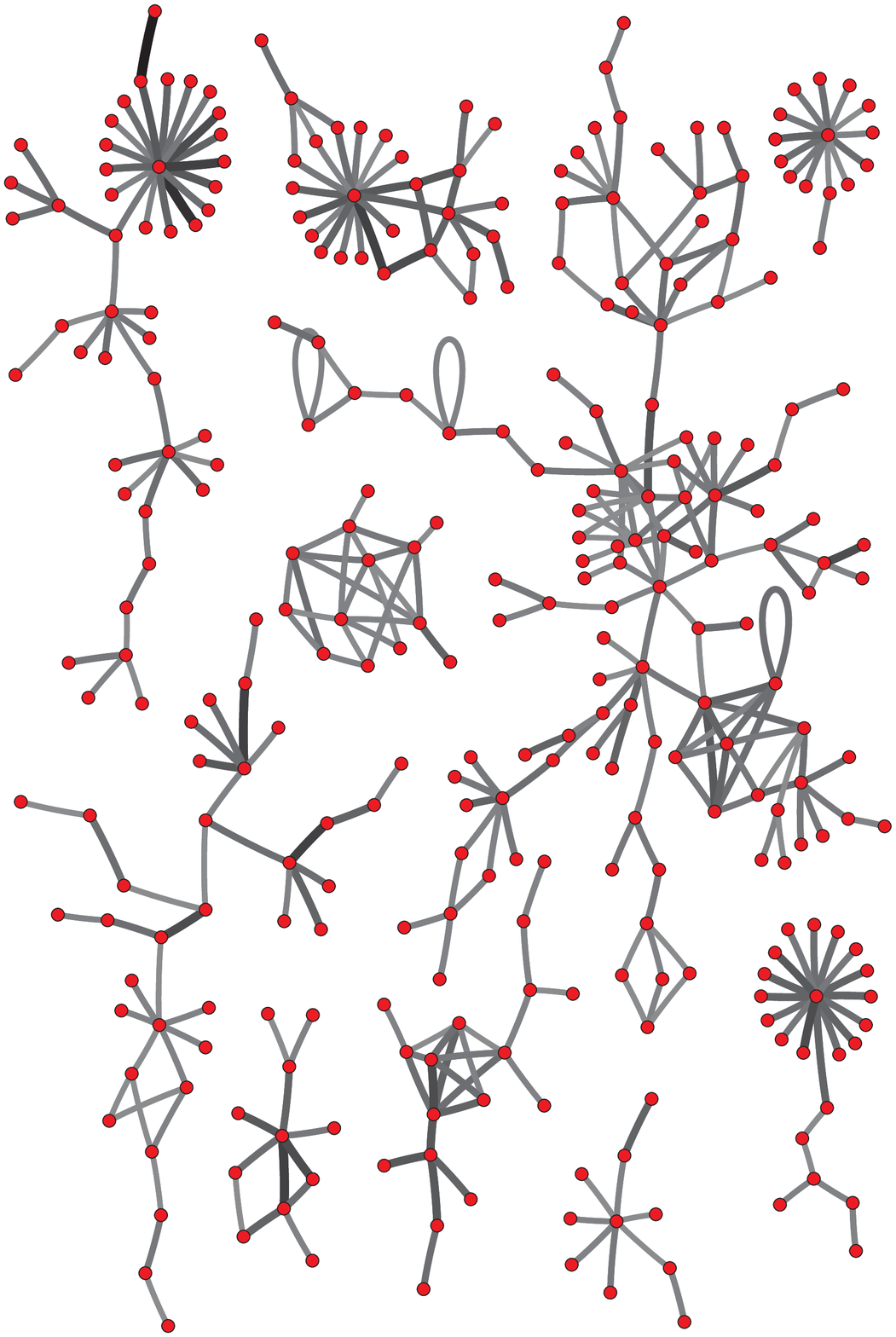}
\includegraphics[width=.3\columnwidth]{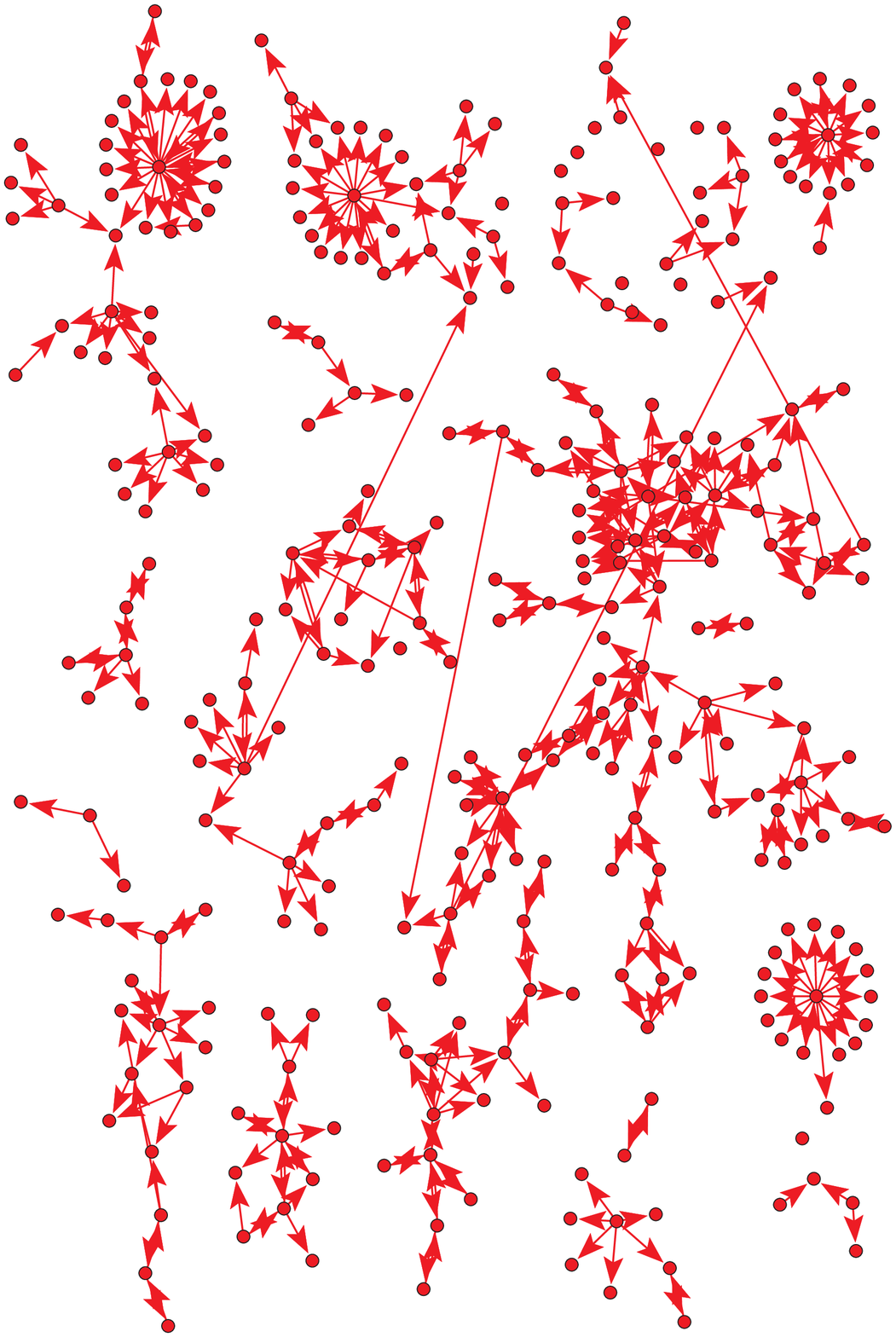}
\caption{\label{fig_3} A) All binding links between a subset of
312 highly abundant proteins.
B) Binding links characterized by high concentration of
heterodimers ($>1000$ molecules/cell).
The level of gray of binding links scales with the logarithm of
concentration of the corresponding heterodimer.
C) Concentration-coupled proteins A $\to$ B with the property that
a twofold increase in the abundance A reduces free concentration
of its immediate binding partner B by 20\% or more. Note that
links roughly coincide with highly abundant dimers shown in the
panel B. Arrows reveal the preferential direction of propagation
of perturbations.
}
\end{figure}

What conditions favor the multi-step propagation of perturbations
along particular channels?
In Fig.~\ref{fig_3}A we show a group of highly abundant proteins
along with all binding interactions between them. Then on
panel B of the same figure we show only those interactions that
according to our LMA calculation give rise to highly abundant
heterodimers (equilibrium concentration $>$1000 per cell). This
breaks the densely interconnected subnetwork drawn in the panel A into
10 mutually isolated clusters. Some of these clusters contain
pronounced linear chains which serve as conduits for propagation
of concentration perturbations. The fact that perturbations indeed
tend to propagate via highly abundant heterodimers is illustrated
in the next panel (Fig.~\ref{fig_3}C) where red arrows correspond
to concentration-coupled nearest neighbors A$\to$B.
Evidently, the edges in panels B and C largely (but not
completely) coincide. Additionally, the panel C defines the
preferred direction of propagation of perturbations from a more
abundant protein to its less abundant binding partners.

To further investigate what causes concentration changes to
propagate along particular channels we took a closer look at eight
three-step chains $A \to A_1 \to A_2 \to B$ with the largest
magnitude of perturbation of the last protein $B$ (twofold
detection threshold following a twofold initial perturbation).
The identification of intermediate proteins $A_1$ and $A_2$ was
made by a simple optimization algorithm searching for the
largest overall magnitude of intermediate perturbations along all
possible paths connecting $A$ and $B$.

\begin{figure*}
\includegraphics[width=0.95\textwidth]{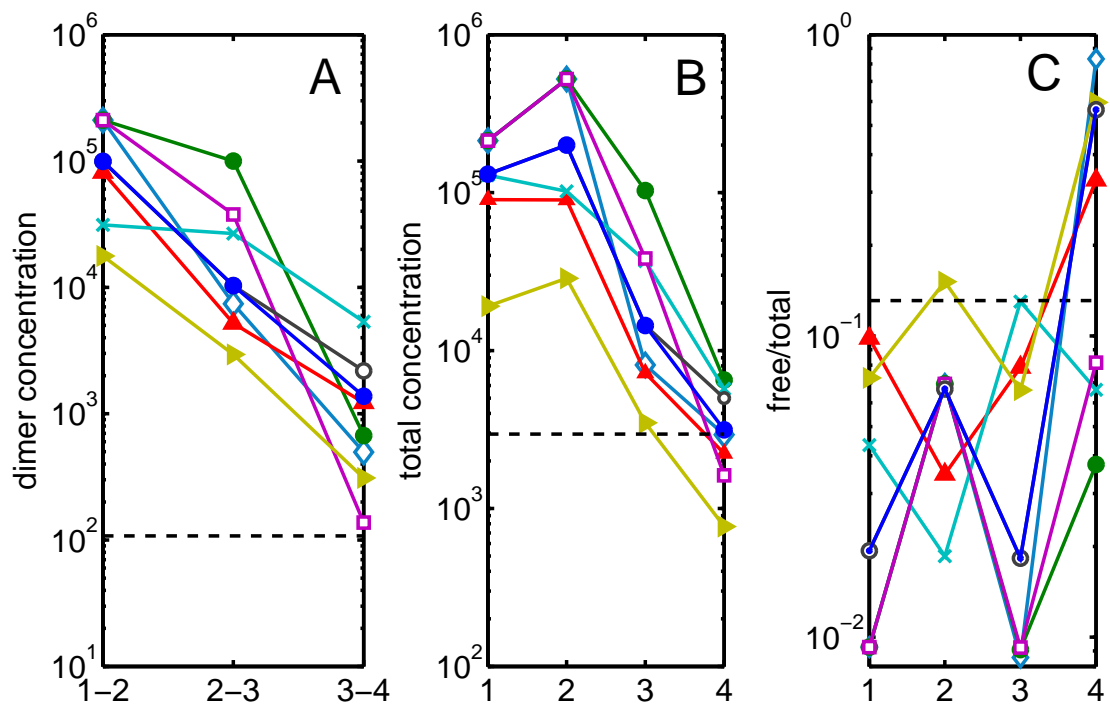}
\caption{ \label{fig_4} Parameters of the eight three-step chains
that exhibit the best transduction of concentration changes:
Heterodimer concentrations $D_{ij}$ (A) for three binding links
along the chain. Total concentrations $C_i$ (B) and free-to-total
concentration ratios $F_i/C_i$ (C) of the four proteins involved
in these chains. Dashed lines correspond to network-wide geometric
averages of the corresponding quantities: $\langle D_{ij} \rangle
\sim 100$ copies/cell, $\langle C_i \rangle \sim 3000$
copies/cell, and $\langle F_i/C_i \rangle = 13$\%.
}
\end{figure*}

Inspection of the parameters of these chains shown in
Fig.~\ref{fig_4} allows one to conjecture that for a successful
transduction of concentration changes, the following conditions
should be satisfied:
\begin{itemize}
\item Heterodimers along the whole path have to be of sufficiently high
concentration $D_{ij}$.
\item Intermediate proteins have to be highly sequestered. That is to say,
in order to reduce buffering effects free-to-total concentration
ratios $F_i/C_i$ should be sufficiently low for all but the last
protein in the chain.
\item Total concentrations $C_i$ should gradually decrease in the direction of
propagation. Thus propagation of perturbations along virtually all
of these long conduits is unidirectional and follows the gradient
of concentration changes (a related concept of a ``gradient
network'' was proposed for technological networks in Ref.
\cite{toroczkai2004jls}).
\item  Free concentrations $F_i$ should alternate
between relatively high and relatively low values
in such a way that free concentrations of proteins at steps 2 and
4 have enough ``room'' to go down. The two apparent exceptions to
this rule visible in Fig.~\ref{fig_4} may be optimized to respond
to a drop (instead of increase) in the level of the first protein.
\end{itemize}
These findings are in agreement with our more detailed numerical
and analytical analysis of propagation of fluctuations presented
in  \cite{prl_us} and illustrated for simple networks in the
Supplementary materials. In  \cite{prl_us} we demonstrated that
the linear response of the LMA equilibrium to {\it small} changes
in protein abundances could be approximately mapped to a current
flow in the resistor network in which heterodimer concentrations
play the role of conductivities (which need to be large for a good
transmission) while high $F_i/C_i$ ratios result in the net loss
of the perturbation ``current'' on such nodes and thus need
to be minimized.

\section*{Discussion}
\subsection*{Robustness with respect to assignment of dissociation constants.}
It has been often conjectured that the qualitative dynamical
properties of
biological networks are to a large extent determined by their
topology rather than by quantitative parameters of individual
interactions such as their kinetic or equilibrium constants (for a
classic success story see e.g. \cite{vondassow2000spn}). Our
results generally support this conjecture, yet go one step
further:  we observe that the response of reversible
protein-protein binding networks to large changes in
concentrations strongly depends not only on topology but also on
abundances of participating proteins. Indeed, perturbations tend
to preferentially propagate via paths in the network in which
abundances of intermediate proteins monotonically decrease along
the path (see Fig. ~\ref{fig_3}). Thus by varying protein
abundances while strictly preserving the topology of the
underlying network, one can select different conduits for
propagation of perturbations.

\begin{figure}[ht]
\includegraphics[width=.75\columnwidth]{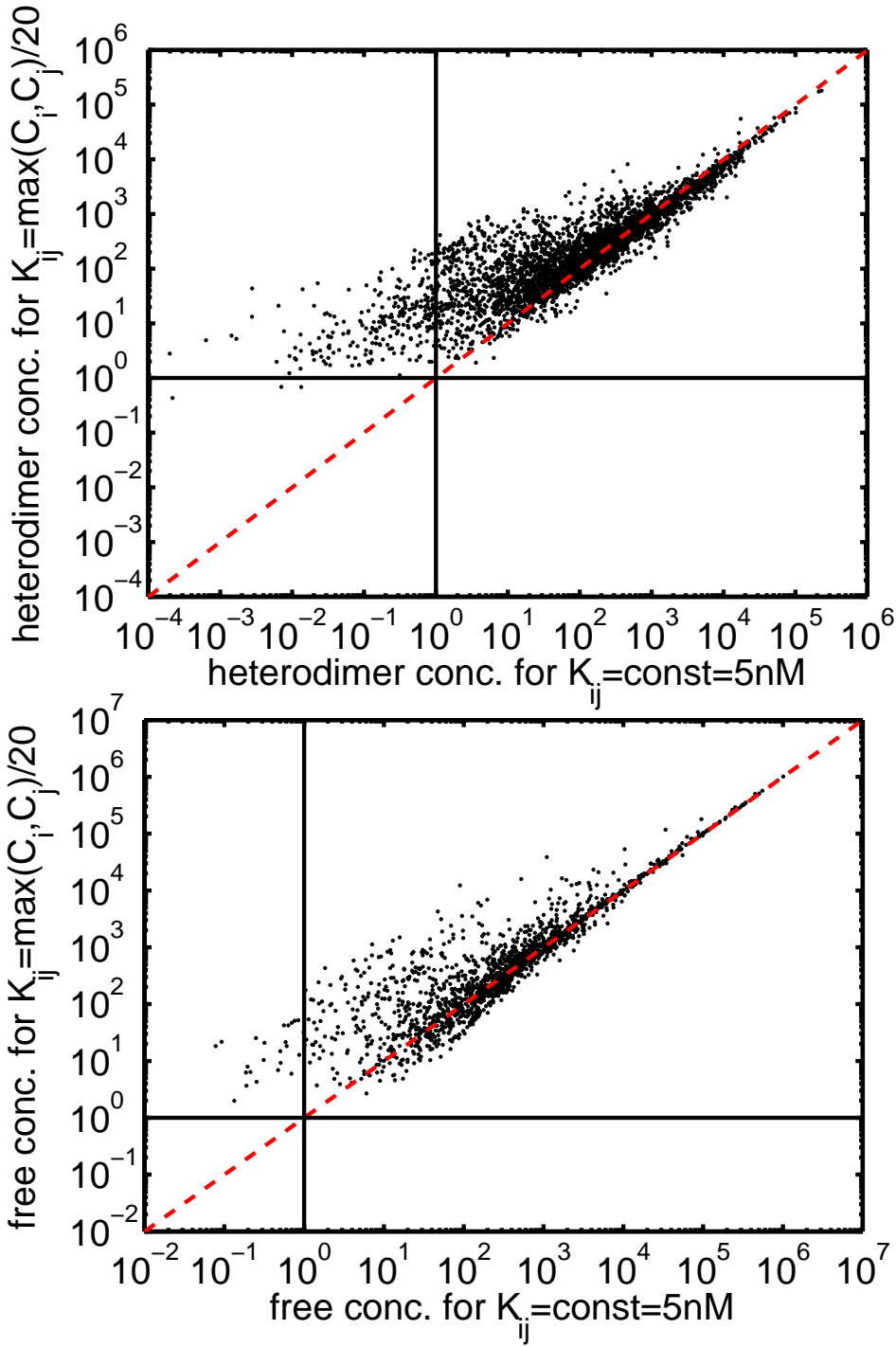}
\caption{
\label{fig5}
The scatter plot of 4185 bound
concentrations $D_{ij}$ (panel A) and 1740 free
concentrations $F_i$ (panel B) calculated for two different
assignments of dissociation constants to links in the PPI network.
The x-axis was computed for the homogeneous assignment
$K_{ij}=\mathrm{const}=5$nM , while the y-axis was computed for the
heterogeneous assignment $K_{ij}=\max(C_i,C_j)/20$
with the same average strength.
The dashed lines along the diagonals are drawn at $x=y$, while the
horizontal and vertical solid lines denote the concentration of 1 molecule/cell.
Note that equilibrium concentrations in the upper part of their
range (e.g. above 1000 molecules/cell) are nearly independent of
the choice of $K_{ij}$. Also, our choice of heterogeneous
assignment nearly eliminates free or bound concentrations
in a biologically unreasonable range $<$1 molecules/cell
}
\end{figure}

On the other hand our results indicate that these conduits are to
a certain degree insensitive to the choice of dissociation
constants. In particular, we found (see Fig.\ref{fig5}) that equilibrium
concentrations of dimers and the remaining free (unbound)
concentrations of individual proteins calculated for two different
$K_{ij}$ assignments ($K_{ij}=\mathrm{const}=5$nM and
$K_{ij}=\max(C_i,C_j)/20$ with the inverse mean of 5nM) had a high
Spearman rank correlation coefficient of 0.89 and even higher
linear Pearson correlation coefficient of 0.98. The agreement was
especially impressive in the upper part of the range of dimer
concentrations (see Fig. \ref{fig5}). For example, the typical difference
between dimer concentrations above 1000 molecules/cell was
measured to be as low as 40\%. As we demonstrated above
it is exactly these highly abundant heterodimers that form the
backbone for propagation of concentration perturbations.
Thus it should come as no surprise that sets of concentration-coupled
protein pairs observed for different $K_{ij}$ assignments also
have a large ($\sim$ 70-80\%) overlap with each other (see the
supplementary table S3).

Such degree of robustness with respect
quantitative parameters of interactions can be partially explained
by the following observation: proteins whose abundance is higher
than the sum of abundances of all of their binding partners cannot
be fully sequestered into heterodimers for any assignment of
dissociation constants. As we argued above, such proteins with
substantial unbound concentrations considerably dampen the
propagation of perturbations, and thus cannot participate in
highly conductive chains. Another argument in favor of this
apparent robustness is based on extreme heterogeneity of wildtype
protein abundances (in the dataset of Ref.
\cite{ghaemmaghami2003gap} they span 5 orders of magnitude). In
this case concentrations of heterodimers depend more on relative
abundances of two constituent proteins than on the corresponding
dissociation constant (within a certain range).

In a separate numerical control experiment we verified that the
main results of this study are not particularly sensitive to false
positives and false negatives in the network topology inevitably
present even in the best curated large-scale data. The percentage
of concentration-coupled pairs surviving a random removal or
addition of 20\% of links in the network generally ranges between
60\% and 80\% (see supplementary table S2).

\subsection*{Genetic interactions.}
\begin{figure}[!ht]
\includegraphics[width=.95\columnwidth]{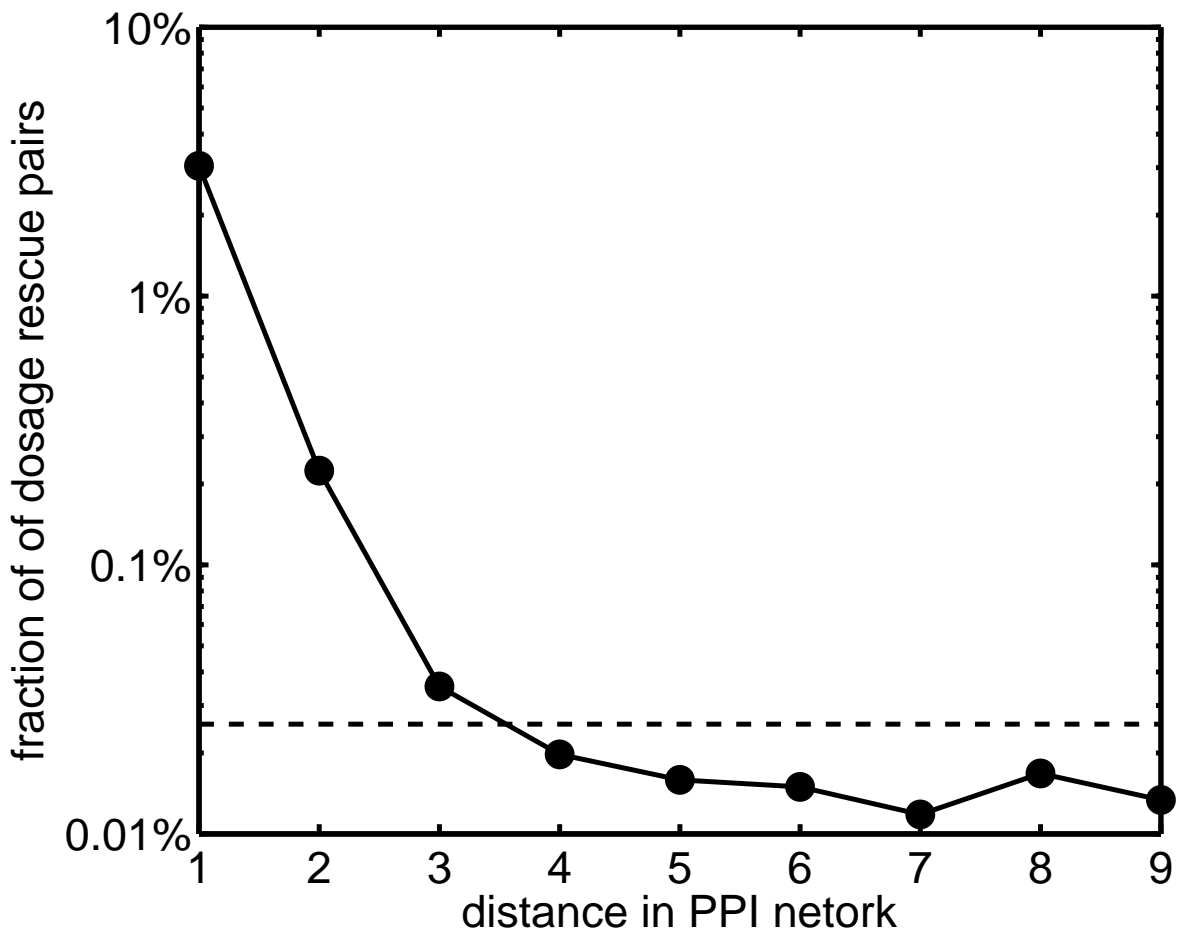}
\caption{
\label{fig6} The fraction of dosage rescue protein
pairs separated by distance $L$ in the PPI network.
Note that pairs at distances 1,2 and 3 are significantly overrepresented over
the background level marked with dashed line ($772/1740^2$) or
visible as a plateau at large distances $L$. The exponential decay
constant at low values of $L$ is consistent with that in Fig.
\ref{fig_2}
}
\end{figure}

The effects of concentration perturbations discussed above could
explain some of the genetic interactions between proteins.
Consider for example a ``dosage rescue'' of a protein $A$ by a
protein $B$, or the correction of an abnormal phenotype caused by
deletion or other type of inactivation of $A$ by overexpression of
$B$. One possible mechanism behind this effect is that the
knockout of $A$ and overexpression of $B$ affect the LMA
equilibrium in opposite directions and to some extent cancel one
another. In order for this mechanism to be applicable (albeit
tentatively), concentrations of both $A$ and $B$ must be
simultaneously coupled (in the sense used throughout this work) to
at least one crucial protein $C$ whose free or bound concentration
has to be maintained at or close to wildtype levels. To assess
this hypothesis, we analyzed the set of 772 dosage rescue pairs
involving proteins from the PPI network used in
this study of 2531 dosage rescue pairs listed in the BIOGRID database
\cite{BIOGRID}. For 136 pairs (or 18\% of all dosage
rescue pairs), we were able to identify one or more putative
``rescued'' protein whose free concentration was considerably (by
$>$20\%) affected by changes in abundances of both  $A$ and $B$
(see supplementary Table S5). This overlap is highly statistically
significant, having the Fisher's exact test p-value around
$10^{-216}$.
Even more convincing evidence that perturbations to the LMA
equilibrium state cause some of genetic interactions is
presented in Figure~\ref{fig6}. It plots the fraction of protein
pairs at distance $L$ from each other in the PPI network that are
known to dosage rescue each other. From this figure one concludes
that proteins separated by distances 1,2, and 3 are significantly
more likely to genetically interact with each other than one
expects by pure chance alone (the expected background level is marked
with a dashed line or better yet
visible as a plateau for large values of  $L$).
Furthermore, the slope of the exponential decay in the fraction of
dosage rescue pairs as a function of $L$ is roughly
consistent with that shown in Fig.~\ref{fig_2} for the fraction of
concentration-coupled pairs.

\subsection*{Possibility of functional signaling and regulation
mediated by multi-step reversible protein interactions.}
Another intriguing possibility raised by our findings is that
multi-step chains of reversible protein-protein bindings might in
principle be involved in meaningful intracellular signaling and
regulation.
There are many well-documented cases in which one-step ``chains''
are used to reversibly deactivate individual proteins by the
virtue of sequestration with their binding partner(s).
An example involving a longer regulatory chain of this type is the
control of activity of condition-specific sigma factors in
bacteria. In its biologically active state, a given sigma factor
is bound to the RNA polymerase complex. Under normal conditions it
is commonly kept in an inactive form by the virtue of a strong
binding with its specific anti-sigma factor (anti-sigma factors
are reviewed in \cite{hughes1998asf}). In several known cases the
concentration of the anti-sigma factor in turn is controlled by
its binding with the specific anti-anti sigma factor
\cite{hughes1998asf}. The existence of such experimentally
confirmed three-step regulatory chains in bacteria hints at the
possibility that at least some of the longer conduits we detected
in yeast could be used in a similar way.

\subsection*{Application to microarray data analysis.}
In order to unequivocally detect cascading perturbations, in our
simulations we always modified the total concentration of just one
protein at at time. In more realistic situations, expression
levels of a whole cluster of genes change, for example, in
response to a shift in environmental conditions. Our general
methods could be easily extended to incorporate this scenario.
With the caveat that changes in expression levels of genes reflect
changes in overall abundances of corresponding proteins, our algorithm allows
one to calculate the impact of an external or internal stimulus measured
in a microarray on free and bound concentrations of all proteins
in the cell. Including such indirectly perturbed targets
could considerably extend the list of proteins affected by a
given shift in environmental conditions. Simultaneous shifts in expression levels of several
genes may amplify changes of free concentrations of some proteins
and/or mutually inhibit changes of others.

\subsection*{Effects of intracellular noise.}
\begin{figure}[h]
\includegraphics[width=.95\textwidth]{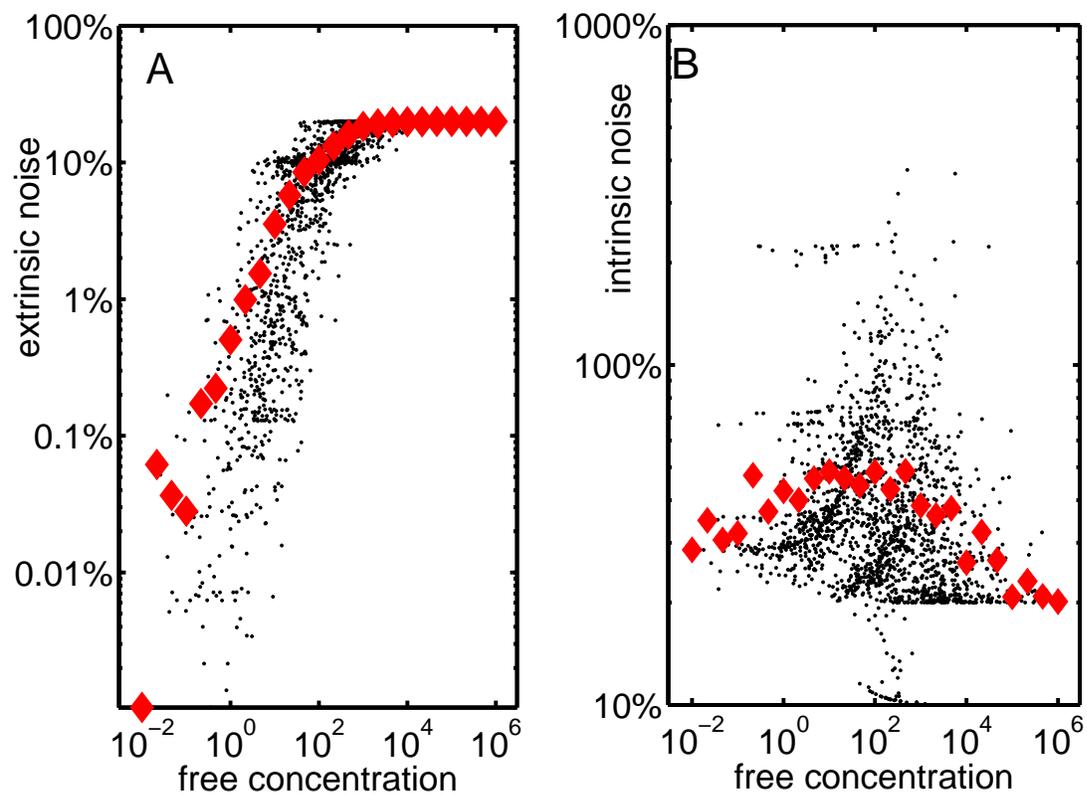}
\caption{
\label{figS4}
The magnitude of extrinsic (panel A) and intrinsic noise
in free concentrations $F_i$ of proteins when their total
concentrations $C_i$ fluctuate by 20\%.
In this plot we used $K_{ij}=\mathrm{const}=1$nM.
One can see that while
the extrinsic noise is suppressed in the low concentrations region,
the intrinsic one is uniformly high and reaches as much as $>$300\%
in the mid-$F_i$ range.
}
\end{figure}

Another implication of our findings is for intracellular noise, or
small random changes in total concentrations $C_i$ of a large
number of proteins. The randomness, smaller magnitude, and the
sheer number of the involved proteins characterize the differences
between such noise and systematic several-fold changes in the
total concentration of one or several proteins considered above.
Our methods allow one to decompose the experimentally measured \cite{newman2006scp}
noise in total abundances
of proteins into biologically meaningful components (free
concentrations and bound concentrations within individual protein
complexes). Given a fairly small magnitude of fluctuations in
protein abundances (on average around 20\% \cite{newman2006scp}),
one could safely employ a computationally-efficient linear
response algorithm (see \cite{prl_us}). Several recent studies
\cite{elowitz2002sge}, \cite{raser2005nge}, \cite{newman2006scp}
distinguish between the so-called extrinsic and intrinsic noise.
The extrinsic noise corresponds to synchronous or correlated
shifts in abundance of multiple proteins which, among other
things, could be attributed to variation in cell sizes and their
overall mRNA and protein production or degradation rates.
Conversely, the intrinsic noise is due to stochastic fluctuations
in production and degradation and thus lacks correlation between
different proteins. We found that extrinsic and intrinsic noise
affect equilibrium concentrations of proteins in profoundly
different ways. In particular, while multiple sources of the
extrinsic noise partially (yet not completely) cancel each other,
intrinsic noise contributions from several sources can sometimes
add up and cause considerable fluctuations in equilibrium free and
bound concentrations of particular proteins
(see Figure \ref{figS4}).

\subsection*{Limitations of the current approach and directions
for further studies.}
In our study we used a number of fundamental approximations and
idealizations including the assumption of spatially uniform
concentrations of proteins, the neglect of temporal
dynamics or, equivalently, the assumption that all concentrations
have sufficient time to reach their equilibrium values,
the continuum approximation neglecting the discrete nature of
proteins and their bound complexes, etc.
Another set of approximations was mostly due to the lack of reliable
large-scale data quantifying these effects. They include not taking into account
the effects of cooperative binding within multi-protein complexes,
using a relatively small number (81) of well curated multi-protein
complexes used in our study (see supplementary
materials), neglecting systematic changes in protein abundances in the course
of the cell cycle, etc. We do not expect these effects to
significantly alter our main qualitative conclusions, namely, the exponential decay of
the amplitude of changes in equilibrium concentrations, the existence of 3-4 step
chains that nevertheless successfully propagate concentration changes,
and the general conditions that enhance or
inhibit such propagation.

In the future we plan to extend our study
of fluctuations in equilibrium concentrations by
incorporating the effects of protein diffusion (non-uniform spatial concentration)
and kinetic effects.
Another interesting avenue for further research is to apply the
concept of ``potential energy landscape'' (for definitions
see \cite{Ao2004} and references therein) to reversible processes governed by the
law of mass action, such as e.g. the equilibrium in protein binding networks.
In the past this concept was applied to processes involving
catalytic, irreversible protein-protein interactions such as e.g.
phosphorylation by kinases or regulation by transcription factors.
In this case it helped to reveal the robustness of regulatory networks in the cell cycle
\cite{Wang2006} and in a simple two-protein toggle switch \cite{Wang2007}.

\section*{Methods}
\subsection*{Source of interaction and concentration data}
The curated PPI network data used in our study
is based on the 2.020 release
of the BIOGRID database \cite{BIOGRID}. We kept only pairs of
physically interacting proteins that were reported in at least two
publications using the following experimental techniques:
Affinity Capture-MS (28172 pairs),
Affinity Capture-RNA (55 pairs),
Affinity Capture-Western (5710 pairs),
Co-crystal Structure (107 pairs),
FRET (43 pairs), Far Western (41 pair),
Two-hybrid (11935 pairs). That left us with 5798
non-redundant interacting pairs. Further restriction for both
proteins to have experimentally measured total
abundance \cite{ghaemmaghami2003gap} narrowed it down
to 4185 distinct interactions among 1740 yeast proteins.

The list of manually curated
yeast protein complexes was obtained from the latest release (May 2006) of
the MIPS CYGD database \cite{MIPS,MIPS_original}.
The
database contains
1205 putative protein complexes 326 of which are not
coming from systemic analysis studies (high-throughput MS experiments). In the spirit of
using only the confirmed PPI data we limited our study to these manually
curated 326 complexes. For 99 of these complexes the MIPS database
lists 3 or more constituent proteins.
After  elimination of proteins with unknown total concentrations we were left
with 81 multi-protein complexes.

Genetic interactions of dosage rescue type were also obtained from
the  BIOGRID database. There are 772 pairs of
dosage rescue interactions among 1740 proteins
participating in our PPI network (the full list
contains 2531 dosage rescue pairs).
\subsection*{Numerical algorithms}.
The numerical algorithm calculating all free concentrations $F_i$ given
the set of total concentrations $C_i$ and the matrix of dissociation constants
$K_{ij}$ was implemented in MATLAB 7.1 and is available for downloading on
\url{http://www.cmth.bnl.gov/~maslov/programs.htm}. It
consists of iterating the Eq.~\ref{eq1} starting with
$F_i=C_i$. Iterations stop once relative change of free concentration
on every node in the course of one iteration step
becomes smaller than $10^{-8}$ which for networks used in our study
takes less than a minute on a desktop computer.
When necessary, multiprotein complexes are
incorporated into this algorithm as described in the Supplementary
Materials.

The effects of large concentration perturbations was calculated by
recalculating free concentrations following a twofold increase in
abundance of a given perturbed protein. The effects of small perturbations such
as those of concentration fluctuations were calculated using the faster linear
response matrix formalism described elsewhere \cite{prl_us}.

%
%
\begin{acknowledgments}
We thank Kim Sneppen for valuable discussions and contributions in
early phases of this project. This work was supported by National
Institute of General Medical Sciences Grant 1 R01 GM068954-01. Work
at Brookhaven National Laboratory was carried out under Division of
Material Science, U.S. Department of Energy Contract DE-AC02-
98CH10886. S.M.'s visit to the Kavli Institute for Theoretical Physics,
where part of this work was accomplished, was supported by National
Science Foundation Grant PHY05-51164.
\end{acknowledgments}




\begin{thebibliography}{10}
\bibitem{linear_response}
As an alternative to this computationally expensive approach we also tried the
  linear response matrix formalism \cite{prl_us} relating small changes in
  $F_j$ to the ones in $C_i$. We found the linear response algorithm to be much
  less computationally expensive, while still providing remarkably good
  approximation to directly computed results even for large changes in protein
  levels.
\end{thebibliography}

\bibliographystyle{pnas}

\end{article}








%
\begin{table}[!ht]
\caption[]{\label{tab1} The number of  concentration-coupled pairs
of yeast proteins separated by network distance $L$.
Numerical simulations (twofold initial perturbation, 20\%
detection threshold) were performed for different assignment of
dissociation constants: $K_{ij}=\max(C_i,C_j)/20$ (column 2),
$K_{ij}=\mathrm{const}=$1nM, 10nM, 0.1$\mu$M,1$\mu$M (columns
3-6). The column 7 lists the total number of protein pairs at
distance $L$.
}
\begin{tabular*}{\hsize}{@{\extracolsep{\fill}}rrrrrrr}
\
%
L &
var. 5nM
& 1nM &
10nM & 0.1$\mu$M & 1$\mu$M & all \cr
\hline
1 & 2003 & 2469    &   1915 & 1184  &    387 &         8168   \cr
2 & 415  & 1195    &    653 &          206  &     71 &        29880 \cr
3 &  15  & 159    &     49 &            8  &      0 & 87772   \cr
4 &  2  &  60    &     19 & 0  &      0 &       228026   \cr
5 & 0  &   3 &      0 &            0  &      0 &       396608 \cr
\hline
\end{tabular*}
\end{table}

\end{document}